\documentclass[useAMS,usenatbib,usegraphicx]{mn2e}
\usepackage[a4paper,margin=2cm]{geometry}
\usepackage{times}
\usepackage{graphicx}
\usepackage{epsfig}
\usepackage{amssymb}
\usepackage{natbib}
\usepackage{journals}
\usepackage[latin1]{inputenc}
\newcommand{\kms} {$\mbox{km s}^{-1}$}

\newcommand{\Msun} {$\mbox{M}_{\sun}$}
\newcommand{\Lsun} {$\mbox{L}_{\sun}$}

\newcommand{\MBH}{{M$_{\rm BH}$}}

\def\spose#1{\hbox to 0pt{#1\hss}}
\def\lta{\mathrel{\spose{\lower 3pt\hbox{$\sim$}}
    \raise 2.0pt\hbox{$<$}}}
\def\gta{\mathrel{\spose{\lower 3pt\hbox{$\sim$}}
    \raise 2.0pt\hbox{$>$}}}
\newdimen\hssize
\newdimen\hdsize
\title[Is the black hole in NGC\,1277 really over-massive?]
{Is the black hole in NGC\,1277 really over-massive?}
\author
[Eric Emsellem]{\parbox{\textwidth}{Eric Emsellem,$^{1,2}$\thanks{E-mail: eric.emsellem@eso.org \texttt{}}
}\vspace{0.4cm}\\ 
\parbox{\textwidth}{$^{1}$European Southern Observatory, Karl-Schwarzschild-Str. 2, 85748 Garching, Germany\\
$^{2}$Universit\'e Lyon 1, Observatoire de Lyon, Centre de Recherche Astrophysique de Lyon \\ \hspace*{0.5cm} and Ecole Normale Sup\'erieure de Lyon, 9 avenue Charles Andr\'e, F-69230 Saint-Genis Laval, France
}}
\begin{document}
\maketitle
%
%
\begin{abstract}
A recent claim has been made by van den Bosch et al. (2012) that the fast-rotator galaxy NGC\,1277 hosts an
over-massive black hole with a mass ($1.7\, 10^{10}$~\Msun) larger than half its (central) stellar spheroid
mass. We revisit this claim here by examining the predictions from simple dynamical realisations based on 
new Multi-Gaussian Expansion models of NGC\,1277, using the same inclination $i=75\degr$, and constant
mass-to-light ratios.  We present realisations which fit well the observed photometry taking into account an
approximation for the extinction due to the central dust ring.  The mass-to-light ratio $M/L$ is fixed
following scaling relations which predict a Salpeter-like IMF for such a luminous early-type fast rotator,
60\% higher than the one of the previously derived best fit model.  A model without a black hole provides a
surprisingly good fit of the observed kinematics outside the unresolved central region, but not, as expected, of the
central dispersion and Gauss-Hermite $h_4$ values. A model with a black hole mass of 5~$10^{9}$~\Msun\ allows
to fit the central dispersion profile, consistently with models of the same mass and $M/L$ in van den Bosch et
al. (2012). It departs from the central $h_4$ values by only about twice the given uncertainty. A slightly
varying $M/L$ or the addition of high velocity stars in the central spatially unresolved region would further
lower the need for a very massive black hole in the central region of NGC\,1277. These results do not, by
themselves, rule out the presence of a presumed over-massive black hole at the centre of NGC\,1277. However, 
they lead us to advocate the use of three-$\sigma$ (as opposed to one-$\sigma$) confidence intervals 
for derived \MBH\ as better, more conservative, guidelines for such studies.
We also caution for the use of ill-defined spheroidal components as an input for scaling
relations, and emphasise the fact that a \MBH\ in the range $2-5\,10^9$~\Msun\ would represent less than 5\%
of the spheroid bulge-like mass of our models and less than $2.5$\% of its total stellar mass. This would make
the black hole in NGC\,1277 consistent or just twice as large as what a recent version of the \MBH-$\sigma$
predicts, well within the observed scatter.  We examine the impact of the presence of an inner bar by running
simulations from the same MGE model but with extreme anisotropies. An inner small (600~pc diameter) bar forms,
and an end-on view does get closer to fitting the central dispersion profile (and fits the $h_3$ amplitude)
without the need for a central dark mass, while adding a black hole of 2.5$\,10^9$~\Msun, in line with the
prediction from scaling relations, allows to fit the dispersion peak and $h_3$ profiles. Both models, however,
still fail to fit the central $h_4$ value (overpredicting the mean velocity).  The claimed large mass of the
presumed black hole therefore mostly relies on the measured positive high central $h_4$ (at high dispersion),
which can be associated with broad wings in the Line-Of-Sight-Velocity Distribution (high velocity stars).
This emphasises the need to go beyond medium resolution long-slit kinematics, with e.g., high resolution
integral-field spectroscopic data.  In the specific case of NGC\,1277, molecular or ionised gas kinematics (if
present) within the central arcsecond (or at large scale) may provide a strong discriminant between these
various models. We finally briefly discuss the fact that 
NGC\,1277 resembles a scaled-up version of e.g., NGC\,4342, another nearly
edge-on fast rotator with a potentially large (but not over-massive) black hole.
\end{abstract}
\clearpage

\section{Introduction\label{sec:intro}}
\begin{figure*}
\centering
\epsfig{file=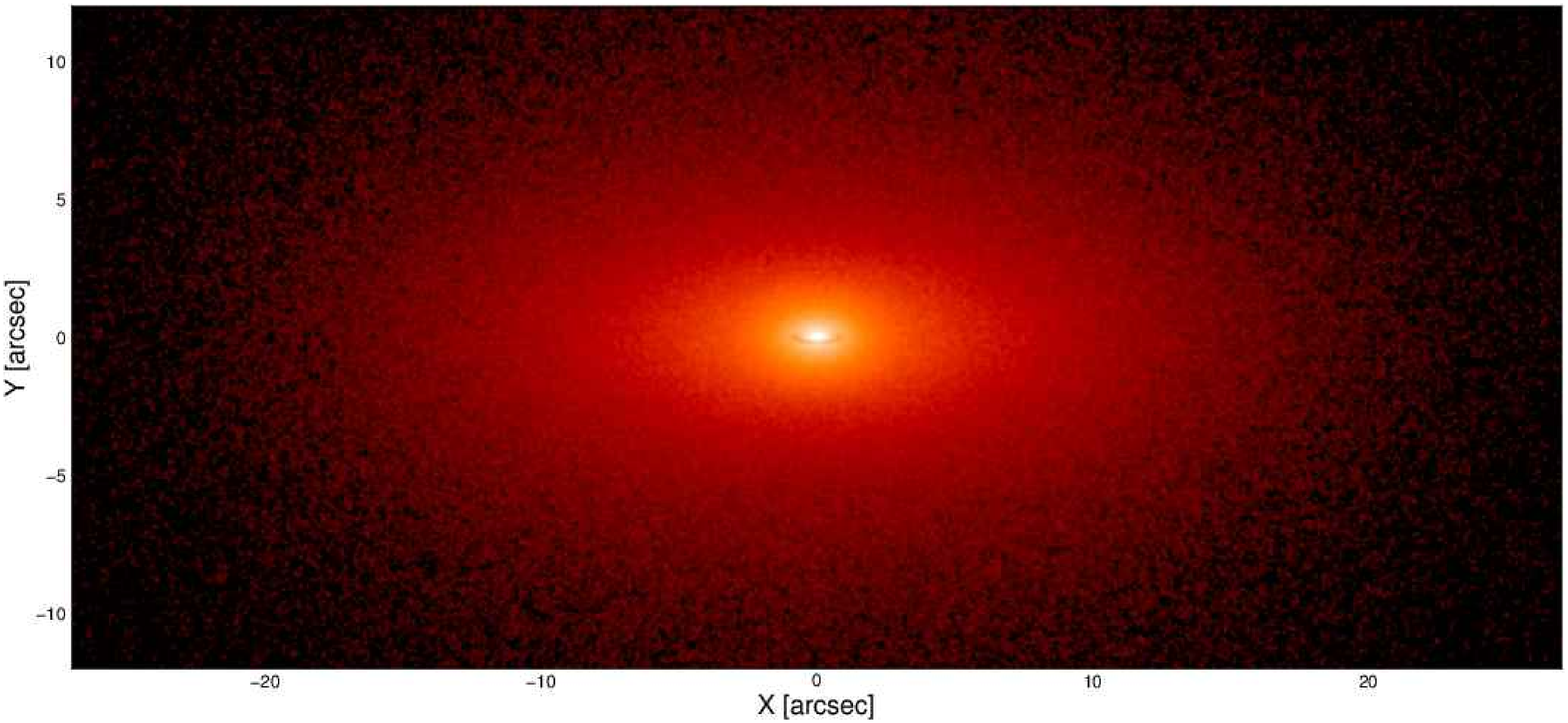, width=17cm}
\epsfig{file=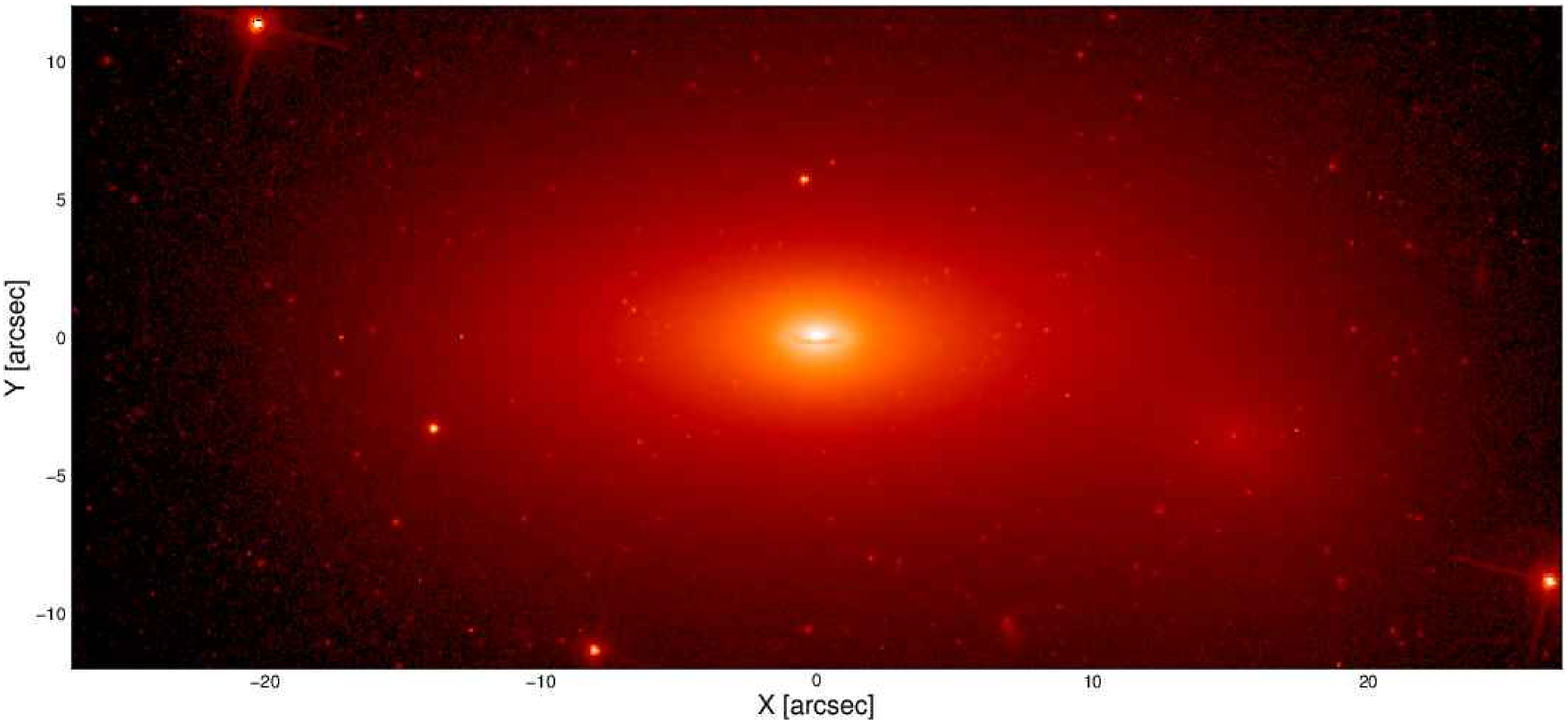, width=17cm}
\caption{Reconstructed image from the MGE realisation presented in this paper (top panel) and the HST/ACS F625W 
data (bottom panel). The MGE model has been projected with an inclination of 75\degr, 
and assumes the presence of a small ring of dust (visible in the central arcsecond) 
which leads to an extinction of about 50\% of the background light. A faint companion is visible in the
bottom right corner of the ACS image. The graininess of the MGE model (top panel) is due to the finite number of
particles included in the realisation (the corresponding analytic MGE model itself being obviously much smoother).}
\label{fig:ImaMGE}
\end{figure*}

A recent claim has been recently made by \cite{vdB+12}
that the fast-rotator galaxy NGC\,1277 hosts an over-massive black hole with a
mass ($17\pm3\times10^9$~\Msun) quoted as larger than half the stellar bulge mass. 
This departs very significantly from predictions
of the so-called \MBH$-\sigma$ relation \citep[see e.g.][]{Tremaine+02}, or the bulge-black hole mass
relation. The best fit model including a relatively small size black hole ($10^8$~M$_{\odot}$) with
respect to its spheroidal component, and using the same mass-to-light ratio, fails to fit e.g., 
the stellar velocity dispersion profile within the central 10\arcsec\ (see their Figure~3). 
This emphasises the fact that a large $17\pm3\times10^9$~\Msun\ black hole
would influence the potential of the galaxy far out, a few hundreds of parsec from its central 
location. If confirmed, the existence of this over-massive black hole (relatively to the host
galaxy) could either be a rare and exotic case, or would actually 
shed new light on the relation between the formation and evolution processes 
of early-type fast rotators and their central dark mass.

In this short paper, we revisit this claim by building simple dynamical models based on the Multi-Gaussian
Expansion technique \citep{Emsellem+94}, also used in the modelling approach adopted by \cite{vdB+12}.
The goal of the present paper is not to perform new fits to the observed dataset. Here, we explore
further what are the main constraints motivating the need for a very large black hole mass,
and wish to examine whether other slight variations in the input assumptions may 
help reconcile the specific case of NGC\,1277 with the above-mentioned scaling relations.
We also examine the potential impact of an inner bar for the observed stellar kinematics
of such a fast rotator \citep[see, e.g.][]{Gerhard88}, and more specifically its potential 
effect on the central stellar velocity dispersion. 
This could result in important diagnostics and in a near future help refute or confirm the claim of \cite{vdB+12}
(hereafter, vdB+12).
In Sect.~\ref{sec:obs} we provide some basic information about 
the observational data and methods we have used in this paper. We then proceed in Sect.~\ref{sec:results}
to present the results coming out of this focused study, and wrap things up in Sect.~\ref{sec:conclusions}.


\section{Observational Data and methods} 
\label{sec:obs}

\subsection{Photometry and stellar kinematics} 

For the present paper, we have used of the same input photometry and stellar kinematics as in \cite{vdB+12},
namely:
\begin{itemize}
    \item Images obtained with the HST/ACS in the course of the GO:10546 programme (PI Fabian) with the
     F550M and F625W filters. We gathered the ACS images from the ST-ECF Hubble Science Data Archive and
     used the PSF as derived from the TinyTim software.
    \item Long-slit stellar kinematics as presented in vdB+12 which consists of a major-axis profile
        from the LRS spectrograph mounted at the Hobby-Eberly Telescope (HET), covering up 
        to about a radius of about 20\arcsec\ for the stellar velocity $V$, velocity dispersion $\sigma$, 
        as well as for the next two Gauss-Hermite parameters $h_3$ and $h_4$ (within the central 7\arcsec\
        only). The spatial resolution is about $1\farcs6$ Full-Width at Half Maximum (FWHM) and is
        parameterised with a double Gaussian model \citep{vdB+12}.
\end{itemize}
The reader is encouraged to consult at the Supplementary information provided by \cite{vdB+12} for
further details.

\subsection{MGE photometric model} 

The dynamical modelling relies on the Multi-Gaussian Expansion (MGE) technique developed by \cite{Emsellem+94}, itself 
partly based on developments by \cite{Bendinelli91} and \cite{Monnet+92}. 
The MGE method is a flexible way of building photometric and dynamical models of
axisymmetric galaxies \citep[see][for a specific treatment in the triaxial case, and deconvolution]{Monnet+92,Emsellem+94} with
complex multi-component structures. A set of two-dimensional Gaussians with various widths and axis ratios
is used to represent the (observed) surface brightness distribution. This MGE representation is 
uniquely deprojected, given viewing angles and assuming that each gaussian component is 
itself deprojected into a  three-dimensional component constant on homothetic ellipsoids. 
The mass distribution can be specified
by attributing a mass-to-light ratio value to each Gaussian, and including additional components (black hole,
or other Gaussians which may serve as a representation for a dark halo or a gas disk).

vdB+12 have built an MGE model for NGC\,1277 using the ACS F550M image and Galfit \citep{Galfit}, the
parameters of which are available in their paper (see their supplemental Table~1). We built a new MGE
model this time using the HST/ACS F625W image: we do not expect that model to differ much from the one 
built by vdB+12 even considering the different implementation for the MGE fitting part itself. 
This new model should help to at least provide some partial assessment on the expected variation in the modelling.
We made various fitting attempts using the {\tt pyMGE} python package (Emsellem et al., in preparation;
see also \cite{Cappellari02} for a robust idl implementation), which led to very similar outputs.
During the fitting process, we masked a few bright stars and a nearby companion, 
and we accounted for the (apparently) roundish early-type galaxy 
NGC\,1278 by first creating an MGE model for that galaxy which was then subtracted from the ACS image. 
We also attempted to mask out the region which suffered 
too much from the dust ring extinction visible within the central arcsecond 
of NGC\,1277 so that our MGE model does not under-predict 
the amount of light at the very centre. The overall approach is similar 
(though not identical) to the procedure followed by vdB+12. 
Since we will subsequently convolve the predicted kinematics with a 
broad point-spread function (PSF, see below), we took into account the overall effect of the ACS PSF by
approximating the corresponding TinyTim prediction with a set of two centred Gaussians.

As mentioned, the dust ring strongly affects the very central region
of the surface brightness distribution (see Figure~\ref{fig:ImaMGE}, or Figure~1 in vdB+12). 
The outer luminosity profile is also not well constrained considering
the limited depth of the ACS images. In order to make sure we do not extend our MGE model significantly beyond
the one built by vdB+12, and to minimise its total flux, we further constrain the two outer Gaussians to be relatively flat
with an axis ratio smaller or equal to 0.7, and to have the same major-axis widths as the ones
in vdB+12. The final model, normalised in flux to the F550M (for comparison) 
is shown in Fig.~\ref{fig:MajMinCut} (see Table~\ref{tab:param}) and compared to the one of vdB+12. Both models are surprisingly similar
although, as expected, our MGE model is slightly flatter in the inner and outer parts. The fact that
we imposed a flattening constraint for the largest Gaussians also prevents our MGE model from showing an abrupt
change in its ellipticity in the outer parts where the observed photometry does not constrain
the model well. Our MGE model has a total $V$-band luminosity 
of about 1.8~10$^{10}$~L$_{\odot}$ as compared to about 
2.0~10$^{10}$~L$_{\odot}$ for the model of vdB+12.
\begin{table}
\caption{Parameters for the reference MGE model, including the maximum amplitude for each 
two-dimensional (projected) Gaussian, its width in arcseconds, and its axis ratio.}
\label{tab:param}
\begin{tabular}{rrrr}
\hline
\# & $\Sigma_0$ & $\sigma_G$ & $q$  \\
& \Lsun.pc$^{-2}$ & \arcsec &  \\
\hline
1 & 116230.070  &     0.0100 &   0.700 \\
2 & 167697.701  &     0.0300 &   0.300 \\
3 &  28108.207  &     0.0700 &   0.700 \\
4 &  11919.362  &     0.1488 &   0.700 \\
5 &   6308.461  &     0.2889 &   0.305 \\
6 &   9483.126  &     0.3208 &   0.700 \\
7 &   6954.700  &     0.6208 &   0.700 \\
8 &   1866.762  &     0.9320 &   0.305 \\
9 &   3551.888  &     1.2155 &   0.695 \\
10 &   1086.092 &     2.6801 &    0.553 \\
11 &    237.533 &     4.4573 &     0.286 \\
12 &    344.707 &     6.5533 &     0.437 \\
13 &     55.810 &     8.9404 &     0.650 \\
14 &     11.970 &    10.5439 &     0.655 \\
15 &      8.385 &    16.9824 &     0.655 \\
\end{tabular}
\end{table}

\begin{figure}
\centering
\epsfig{file=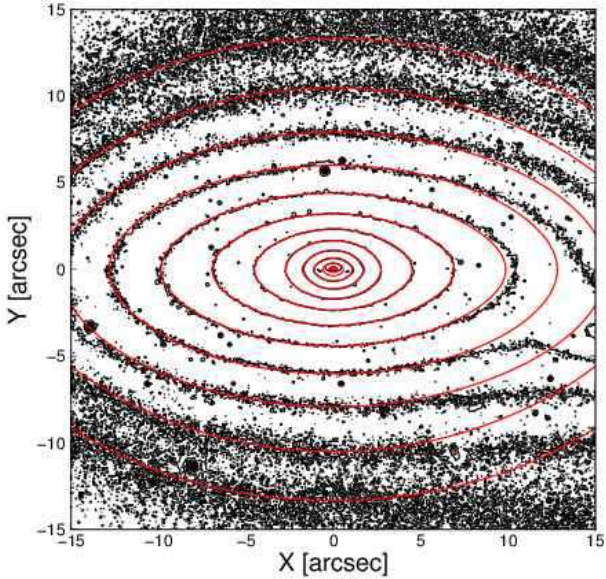, width=\columnwidth}
\caption{Isophotes of our MGE model (red contours) compared to the input HST/ACS F625W image (black contours). 
The influence of the galaxy
neighbour in the South-West quadrant is visible already at 10\arcsec\ along the major-axis of NGC\,1277.}
\label{fig:ContMGE}
\end{figure}
\begin{figure}
\centering
\epsfig{file=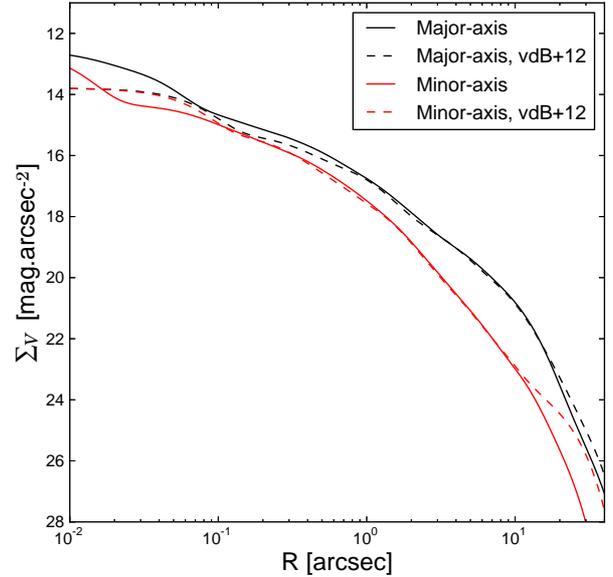, width=\columnwidth}
\caption{Major- and minor-axes photometric surface brightness 
cuts of the MGE models built in the present paper (solid lines), compared to the input MGE
model (dashed lines) from van den Bosch et al. (2012). 
Profiles have been normalised to $V$-band surface brightness profiles for comparison.}
\label{fig:MajMinCut}
\end{figure}

\subsection{MGE dynamical models and comparison with observations} 
\label{sec:dyn}

We built dynamical models from N-body realisations of the MGE axisymmetric models described above, 
using {\tt pyMGE}. This is achieved by deprojecting the MGE models with an inclination 
of 75\degr, assuming a constant stellar mass-to-light
ratio $M/L$ for all Gaussians, and solving Jeans Equations as in \cite{Emsellem+94} and \cite{Cappellari08}, 
constraining the velocity ellipsoid associated with each Gaussian
to have constant axis ratios, namely $\sigma_R / \sigma_z$ and $\sigma_{\theta} /
\sigma_R$ (where $(R, \theta, z)$ are the cylindrical coordinates). We further added a central
point-like dark mass to the gravitational potential, playing thus the role of a black hole, 
but no outer dark component. We kept the freedom to reverse the orbital sense of rotation of stars, as this is 
a way to influence the projected odd velocity moments significantly
(e.g., the projected mean velocity $V$ and $h_3$) without impacting the 
photometry (and mass) distribution. The realisations are compared with observations by first applying
a simple projection, and then taking into account the 
seeing Point Spread Function (PSF), pixel size and nature of the observables (photometry, kinematics via its Gauss-Hermite
moments). For the kinematic quantities, we reconstructed the Line-of-Sight-Velocity Distributions with
velocity bins of 25\kms\ steps, which are subsequently fitted using Gauss-Hermite functions up to 
the fourth degree leading to a measure of $V$, $\sigma$, $h_3$ and $h_4$.
We performed realisations using from $10^5$ up to $4\times10^6$ particles,
and we checked that the total number of particles does not significantly affect the output 
projected quantities.

For comparison with the HST/ACS images, we approximated the observed extinction due to the
presence of a dust (and presumably gas) ring in the central arcsecond by adding a simple "screen" model: the ring itself
is modelled as a circular ring of dust in the equatorial plane (between 0\farcs5 and 0\farcs95) 
and a maximum extinction factor of about 1.7 magnitude. This dust model is not 
meant as a realistic account of the dust extinction, but only as
a way to get a first estimate of the effect of such a ring. This has 
obviously a significant effect on the reproduced HST photometry (see Figure~\ref{fig:ImaMGE}), 
but we find that it does not have a significant
impact on the observed kinematics at the spatial resolution of the HET data.
\begin{figure}
\centering
\epsfig{file=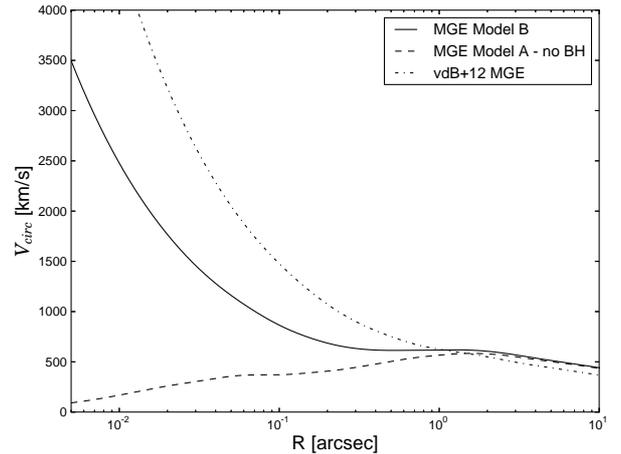, width=\columnwidth}
\caption{Circular velocity profiles for our MGE models with $M/L=10$ and no black hole (Model~A, dashed line),
or a black hole mass of 5~$10^{9}$~M$_{\odot}$ (Model~B, solid line),
and the corresponding profile (dashed-dotted line) for the best fit model of \citet{vdB+12} with $M/L=6.3$ and
a black hole of 17~$10^{9}$~M$_{\odot}$.}
\label{fig:Vcirc}
\end{figure}

As mentioned above, we used a constant mass-to-light ratio $M/L$ as in \cite{vdB+12}. 
We define a reference value, by simply following the 
known trend of more massive early-type galaxies having higher $M/L$ (see Cappellari et al. 2012, 
and references therein, and e.g., Cappellari et al. 2006, Thomas et al. 2011). 
vdB+12 quote an aperture (within one effective radius but excluding
the central arcsecond) stellar velocity dispersion of $\sigma_e \sim 333$~\kms, and the total $V$-band luminosity 
of NGC\,1277 is around $2\,10^{10}$~\Lsun: this means that we expect a $M/L$
consistent with a close to Salpeter-like initial mass function (IMF). The same is true if we 
refer to the mass of the galaxy which is between 1 and 2~10$^{11}$~M$_{\odot}$. 
We therefore adopted a reference value of $M/L_V = 10$,
about 60\% larger than the value of 6.3 for
the best fit model in vdB+12. The circular velocity profiles in the equatorial plane 
for both this model (with and without a black hole) and the best fit model presented in vdB+12 
(including an "over-massive" black hole of
17~$10^9$~M$_{\odot}$) are shown in Fig.~\ref{fig:Vcirc}. We should note that what drives
the combined amplitudes of the stellar kinematics in the outer part (e.g., the second order velocity
moment $V^2+\sigma^2$) is the total mass, inclusive of the stars and dark matter if present. Having 
$M/L$ set to a Salpeter-like IMF without a dark halo, could be thus roughly equivalent to having
a lower value of the stellar $M/L$ but adding dark matter (which would require to additionally define 
its relative scale, mass normalisation, and shape): this is illustrated in the shift towards
higher $M/L$ of the chi-square contours obtained by vdB+12 (Figures 1 and 3 of their Supplementary material)
when removing the dark halo in the models. For a galaxy of the mass of NGC~1277, we would expect
a dark matter fraction of about 15\% \citep{Cappellari+12} within one effective radius (~5\arcsec), and
probably dominating the mass budget beyond $\sim 15$\arcsec. Since we 
focus here on simple models, and do not try to actually obtain best fits to the observed
kinematics, we voluntarily exclude the use of an additional dark matter component in the outer part 
(a dark halo) or a varying $M/L$ profile.

We used the {\tt Gadget2} N-body code developed by \cite{Gadget2} to both test the stability of our MGE
N-body realisations, as well as the impact of the formation of a bar. For the latter, we imposed
a more extreme anisotropy with a smaller tangential dispersion $\sigma_{\theta}$ : this basically forces the
model to start with dynamically cold initial conditions and is an efficient way to enforce the (artificial)
triggering for the formation of a non-axisymmetric system (such as a bar).


\section{Results} 
\label{sec:results}

\begin{figure}
\centering
\epsfig{file=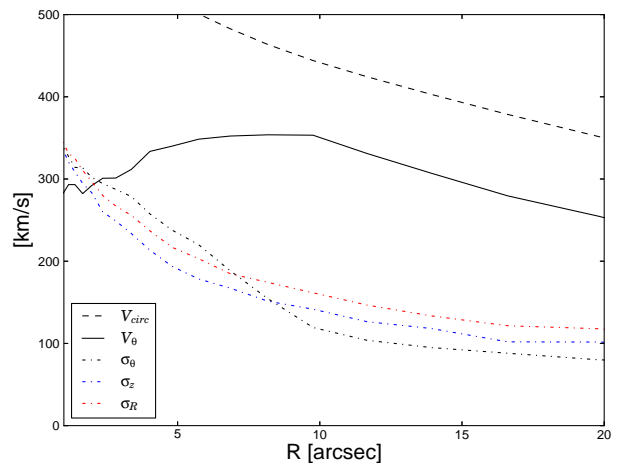, width=\columnwidth}
\caption{Velocity moments in the equatorial plane of the MGE Model~B presented in this paper:
the circular velocity (dashed line), mean tangential velocity (solid line) and 
velocity dispersion profiles (dotted-dashed lines; radial, azimuthal and vertical dispersion, respectively
as red, black, and blue lines).}
\label{fig:Sigma}
\end{figure}

\begin{figure}
\centering
\epsfig{file=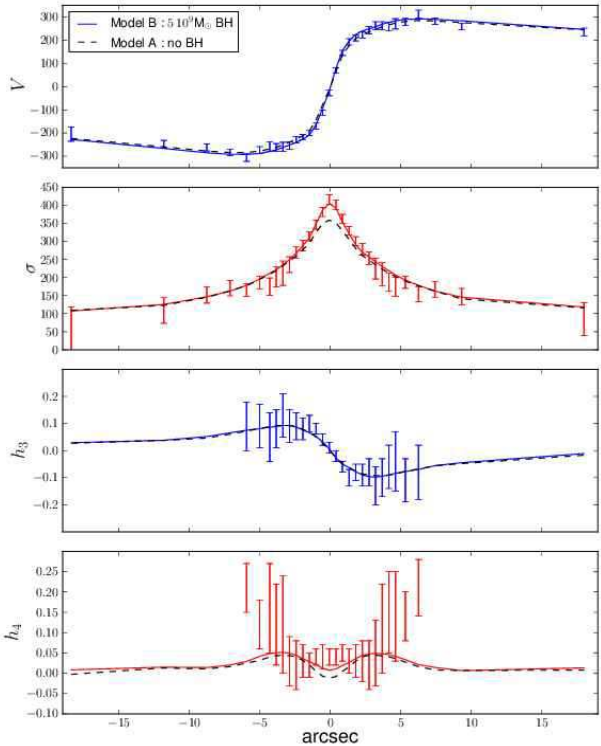, width=\columnwidth}
\caption{Kinematics from the HET long-slit data \citep{vdB+12}, compared with the two models discussed in the
present paper (Model~B with a black hole mass of $5\,10^{9}$~\Msun, as solid lines, and Model~A without a black hole
as dashed lines). 
This figure corresponds to Figure~3 of \citet{vdB+12}.}
\label{fig:kin}
\end{figure}

For our reference (axisymmetric) models, we impose a simple anisotropy profile with $\sigma_R / \sigma_z \sim 1.15$ and  $\sigma_R / \sigma_{\theta} \sim 1.4$. The first ratio corresponds to $\beta_{Rz} = 1 - \sigma_z^2 / \sigma_R^2 \sim 0.25 $, which is incidently consistent with the general trend found by \cite{Cappellari+07} where $\beta_{Rz} \sim 0.7 \times \epsilon_{intr}$ with $\epsilon_{intr}$ being the intrinsic ellipticity, assuming an ellipticity of about 0.3 for NGC\,1277. The relatively high tangential anisotropy of the model leads to a high amplitude for the azimuthal rotation, making it a clear fast rotator
(Fig.~\ref{fig:Sigma}). We built realisations with various black hole masses using this reference model: in Fig.~\ref{fig:kin} we illustrate two models, one without a black hole (\MBH$=0$, Model~A), and one with a black hole (\MBH$=5\,10^9$~\Msun, Model~B) twice as big as the one predicted by the M$_{BH}$-$\sigma$ relation recently provided by \cite{McConnell+11}. 
Both models are stable when run with {\tt Gadget2} (pure N-body run with $4\,10^6$ particles 
with various tested softenings between 10 and 50~pc).
Model~B corresponds to a model which departs from the best fit model of vdB+12 at more than the 3-$\sigma$ level (without a
dark halo, see their Supplemental Figure~3), while we obviously expect Model~A (without a black hole) to 
be a very poor fit of the observed stellar kinematics.
In both cases we reversed the sense of rotation for about 12\% of the stars for the rounder inner gaussian components, as to follow the observed skewed LOSVD ($h_3$ being significantly non-zero): this basically
increases the azimuthal dispersion within the central 10\arcsec\ or so (see Fig.~\ref{fig:Sigma}). 

These models were not designed as actual fits for the observed stellar kinematics,
but can reproduce the observed kinematics profiles rather well 
over most of the available radial ranges, and down to about 1\arcsec. This may look surprising considering the 
very poor fit of the model with a low mass ($10^8$~\Msun) black hole presented 
by vdB+12 in their Figure~3. However, that fit was constrained to have the same mass-to-light
ratio than the best fit model, with a low value of 6.3, 
thus being very far from the best fit model without (or with a low-mass)  
black hole. We also compared the best fit model with \MBH$=10^8$~\Msun\ (Remco van den Bosch, priv. comm.)
with our Model~A at the same input $M/L$: our model is a significantly better fit to the kinematics, 
but this is mostly in the outer parts while in the inner region, the two models are similar (the Schwarzschild
technique provides a slightly but not significantly better fit as expected). 
The reason for this is certainly associated with the rounder outer Gaussians included in vdB+12 MGE model,
which may render more difficult the fit of the (high $V/\sigma$) kinematics at large radii. It, however,
does not invalidate the analysis in the central part. As expected, Model~A (model without a black hole) 
does not fit the central dispersion value and has a slightly too shallow central velocity gradient. Model~A,
however, represents already quite a remarkable fit of the outer region down to the central 2\arcsec.
The addition of a 5~$10^{9}$~\Msun\ black hole (Model~B) solves these two above-mentioned discrepancies. 
The $h_3$ profile is fitted rather well independently from the mass of the black hole which is used: 
this is mostly the consequence of the presence of about 10\% of counter-rotating stars which skew the observed LOSVD. 
The only significant discrepancy between Model~B
and the observed stellar kinematics lies in the $h_4$ profile at the very centre, where the predicted values
are slightly two low (0.0-0.02 instead of 0.04). That difference is, however, rather small, and only 
twice the given uncertainty on these parameters. The model of vdB+12 which corresponds to 
the $M/L$ and \MBH\ values of our Model~B provides a slightly better fit of the central region,
while it provides a too low predicted $V/\sigma$ in the outer part, again very probably 
due to its rounder Gaussians (a situation similar to the one described for Model~A).

A high $h_4$ value is associated with high velocity wings of the LOSVD in the central arcsecond.
A spurious $h_4$ value could for example originate in a so-called template mismatch
(a mismatch between the galaxy spectra at these positions and the stellar template used
to extract the kinematic parameters). However, if confirmed, such a high $h_4$ value (associated with a high
central dispersion) would, for the central LOSVDs of NGC\,1277, actually correspond to a few $10^8$~\Msun\ stars at
velocities between 900 and 1100~\kms: these are missing in Model~B. To remove that discrepancy, we could
artificially add stars with high angular momentum (circular orbits) within 0\farcs1 ($\sim 35$~pc), 
where the circular velocity is in the right range (see Fig.~\ref{fig:Vcirc}). This would, however, 
lead to a significant discrepancy in the photometry (assuming a constant $M/L$).
A more massive black hole (e.g., $17\,10^9$~\Msun, as in vdB+12) naturally predicts the required 
high velocity wings by increasing the volume within which such velocities are reached.
This central $h_4$ value is therefore a strong signature of large mass excess in the region unresolved
by the HET spectroscopy, and is (as expected) a strong discriminant between models with various black hole masses.

\begin{figure}
\centering
\epsfig{file=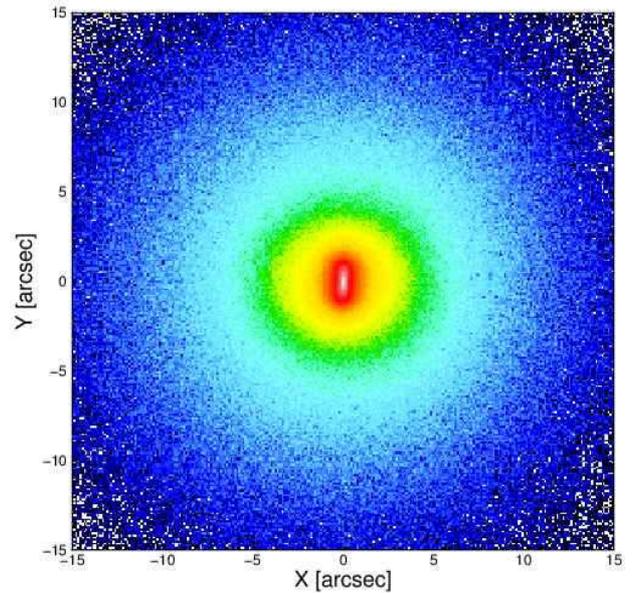, width=\columnwidth}
\caption{Face-on view of the central region of Model~C after 160~Myr of evolution: the mini-bar is clearly
visible here with a diameter of about 600~pc.}
\label{fig:imabar}
\end{figure}

\begin{figure}
\centering
\epsfig{file=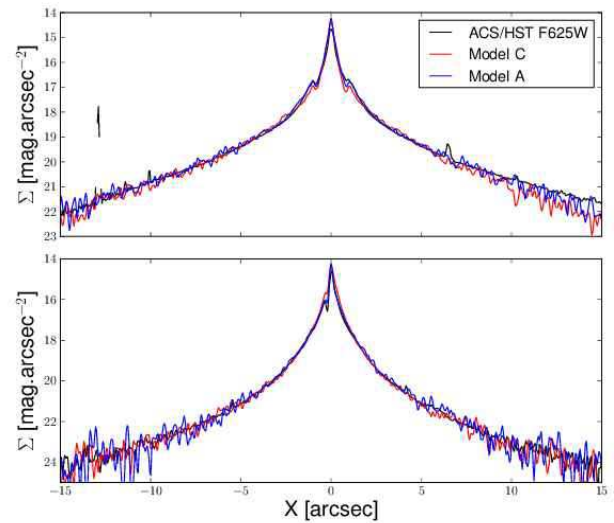, width=\columnwidth}
\caption{Cuts along the major and minor-axes of Model~C after projecting the bar end-on (red profiles)
and compared with the ACS/HST F625W surface brightness profiles.}
\label{fig:cutbar}
\end{figure}

We also examined the impact of the presence from a potentially small bar in these
models by running Gadget2 N-body simulations with more extreme anisotropies, this time setting up
the black hole mass to zero (Model~C) or to a value of $2.5\,10^9$~\Msun, consistent with scaling relations
(Model~D). As mentioned in Sect.~\ref{sec:dyn},
we force a small tangential anisotropy dispersion by setting a high value of $\sigma_R / \sigma_{\theta}$,
namely $2.0$ here. With such initial conditions a small bar quickly 
forms after a few dynamical times in the simulation, and stays as long the simulation was run
(200~Myr). The extent of the bar is small, with a diameter of about 600~pc, as illustrated 
in the face-on view in Fig.~\ref{fig:imabar}.
When the bar is edge-on (and the model inclined at 75\degr), the photometry is clearly not consistent with the observed
photometry. With the bar seen end-on, the fit to the photometry is rather good as shown in Fig.~\ref{fig:cutbar},
which is a bit surprising considering the significant mass redistribution triggered by the formation of the bar.

\begin{figure}
\centering
\epsfig{file=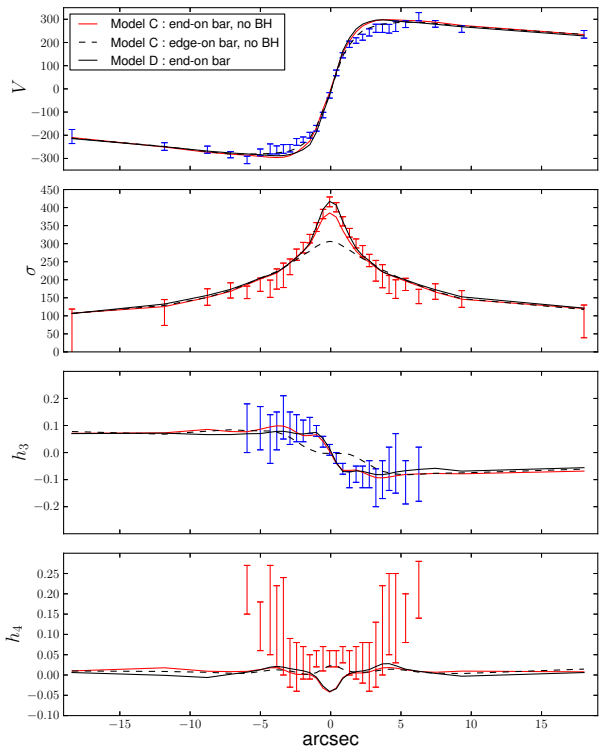, width=\columnwidth}
\caption{Comparison of the predicted stellar kinematics of Model~C with the bar end-on (red lines) and
edge-on (dashed black lines), and the end-on view of Model~D (black solid lines) 
as compared with the observed kinematics from \citet[][errors bars]{vdB+12}. Note that Model~C has no black hole,
and Model~D one with \MBH$=2.5\,10^9$~\Msun.}
\label{fig:barkin}
\end{figure}

We then compare the end-on and edge-on views of these two models (Model~C and D) 
with the observed stellar kinematics of \cite{vdB+12}
in Fig.~\ref{fig:barkin}. Using the same mass-to-light ratio ${M/L}_{V} = 10$, 
we get a reasonable fit of the global profiles even though this model is now, as expected, 
slightly under-predicting the velocity dispersion and over-predicting the mean velocity.
An end-on view of the bar leads to an increase in the central dispersion peak which gets close to
the observed values (385 instead of 415~\kms) for Model~C and fits the dispersion peak for Model~D.
It is only a relative surprise considering that an end-on bar
has already been discussed more than twenty years ago as a potential way to maximise the observed velocities 
of stars \citep{Gerhard88}. It is also interesting to note that,
in both models, the end-on bar naturally produces a significant kink in the $h_3$ 
profile which follows rather closely the observed radial profile without the need for counter-rotating
stars as in Model~A and B. Ironically, the fit to the $h_4$ profile is 
reasonably good when the bar is near edge-on, but with the predicted central $h_4$ 
becoming negative for an end-on view. Again, the central $h_4$ seems to be the most 
constraining observed parameter for the presumed central dark mass.
Obtaining a fit using such an evolved (barred) model is clearly beyond the scope of this paper. 
Results from Model~C and D seem to indicate, however, that the presence of such a non-axisymmetric system may
eventually lower the required central dark mass (even though the coexistence of a central dark mass
and an inner 600~pc bar would have to be explained). It is worth noting that 
the bar which forms in these models would be located just inside the observed dust ring:
this is very probably just linked with the fact that the circular velocity of Model~A 
(axisymmetric model, and no black hole) peaks near a radius of 1\arcsec. 

\section{Discussion and conclusions}
\label{sec:conclusions}
\begin{figure}
\centering
\epsfig{file=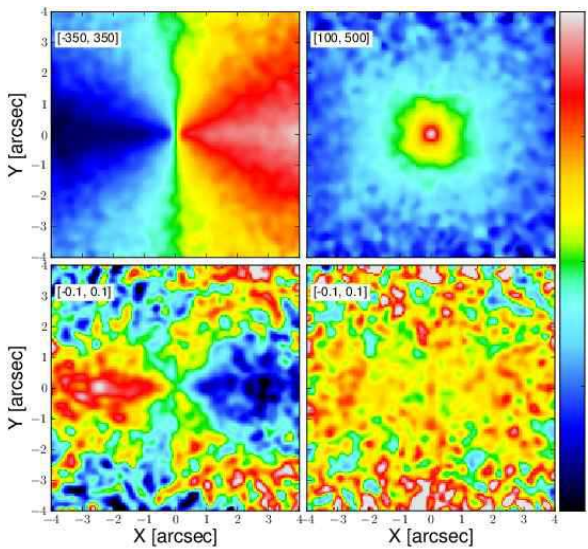, width=\columnwidth}
\caption{Predictions from Model~B for the stellar mean velocity $V$ (top left), velocity dispersion $\sigma$ (top right), 
$h_3$ and $h_4$ fields (bottom left and right, respectively) at a resolution of $0\farcs3$~FWHM. $V$ reaches
about 300~\kms, the dispersion peaks at around 500~\kms, $h_3$ reaches between 0.075 and 0.1 and $h_4$ is
fairly constant with values around 0.03-0.04. Cuts (min/max) are : $\pm 350$~\kms for $V$, $100 / 500$~\kms for $\sigma$,
and $\pm 0.1$ for the $h_3$ and $h_4$ maps. The inclination angle is 75\degr.}
\label{fig:ifu}
\end{figure}
\begin{figure}
\centering
\epsfig{file=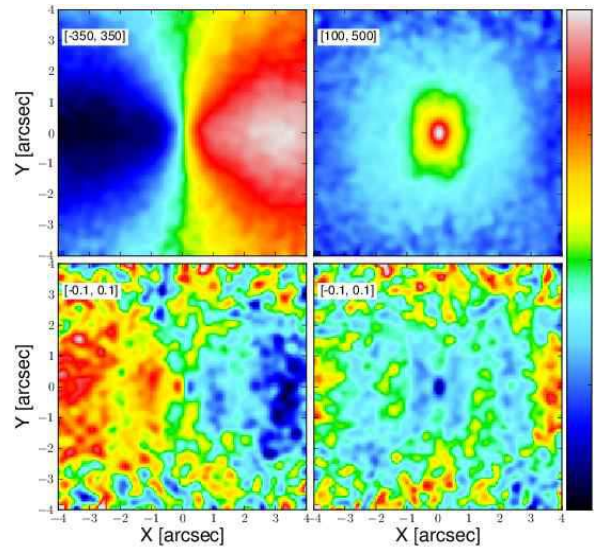, width=\columnwidth}
\caption{As in Fig.~\ref{fig:ifu} but for the (end-on) barred Model D. 
The predicted central dispersion again reaches $\sim 500$~\kms, not making
it a strong discriminant between the models with or without such bars. However,
the elongated shape of the dispersion peak could help constraining the black hole mass,
as well as the two-dimensional shape and amplitude of the high order moment maps.
In the present plot, $h_3$ goes up along the major-axis at 0.06 close to the centre 
(with a local minimum around 2\arcsec\ away along the major-axis), 
and $h_4$ is close to zero everywhere or slightly negative.  The inclination angle is 75\degr.}
\label{fig:barifu}
\end{figure}

In this short paper, we revisit the recent claim for the presence of an over-massive black hole
in NGC\,1277 by van den Bosch et al. (2012). We build new MGE models which fit the photometry, 
using an approximation for the extinction of the observed dust ring, and assuming a fixed $M/L = 10$
consistent with a Salpeter IMF, and a nearly constant anisotropy. 

We show that a model without a black hole does indeed not fit the central dispersion and
$h_4$ values, as expected, but still provides a rather good global fit of the observed long-slit kinematics.
The corresponding models (with similar $M/L$ and \MBH\ values) of vdB+12 have some difficulty to
follow the high $V/\sigma$ values observed in the outer parts: this may
be partly linked with the intrinsic flattening of the outer Gaussians, and the
flatter external components of our MGE models seem to significantly improve the situation
in this context. The model with a very low \MBH\ value presented in Figure~3 of vdB+12 corresponds to the (low) 
mass-to-light ratio of their overall best fit model. It is clearly a very bad fit to the observed
kinematics over nearly the entire radial range. However, this may be somewhat misleading as it 
naturally emphasises the tension between models with and without a central dark mass, not
following the ridge of minimum chi-squared in the $M/L$-\MBH\ diagrams in vdB+12.
We find it remarkable that a model with a constant $M/L$ following observed scaling relations,
simple anisotropy profiles, and no black hole provides a rather reasonable fit to 
the large-scale (major-axis) stellar kinematics of the galaxy except for the central arcsecond. 
The addition of a 5~$10^9$~\Msun\ central dark mass removes most of the discrepancy with the 
central stellar velocity dispersion value, and has a predicted
$h_4$ value within two times the quoted uncertainty on that parameter.

We must emphasise again that the models presented in this paper are not actual fits to the data, 
but only MGE models which fit the photometry and include simple assumptions for its mass-to-light ratio
and anisotropy: they can therefore not lead to a direct statement with respect to the 
"best-fit" value for \MBH, and the value claimed by \cite{vdB+12}. By contruction, the Schwarzschild
techique does find the best fit value given a set of assumptions (e.g., MGE model).
Still, there are reasons to doubt the strict validity of the confidence interval
for the \MBH\ value derived by Schwarzschild modelling. Some doubt may be raised when we witness
the difficulty of the $M/L=10$ models of vdB+12 to fit the outer high $V/\sigma$ values
very probably due to the roundish outer Gaussians of their MGE models. In that context, the role
of the black hole becomes very prominent as to allow the flexibility (with a lower $M/L$ value) 
to fit the entire radial range. The addition of dark matter, if in a spherical halo, should not impact
much on these issues, except by lowering the need for a high $M/L$ to fit the overall amplitude
of the kinematics (e.g., $V^2 + \sigma^2$) in the outer part. 
Looking at the simple parameterisation of Model~A (constant $M/L$, no black hole) and the overall quality of the fit it provides
outside the central 2\arcsec, it is tempting to use an argument in the spirit of Occam's razor to discard a
very high \MBH\ for NGC~1277. Since chi-square contours should provide (as presented in vdB+12) 
an objective assessment of the quality of the fits, but do not include the systematic changes 
which could be triggered by a change in the MGE model itself
(via e.g. different Gaussians, or a slightly varying $M/L$), we believe that 
it solely means that we should be more conservative
and advocate to quote uncertainties corresponding to 3-$\sigma$ confidence levels 
instead (which would nearly include Model~B),
namely the best fit value of $17\,10^9$~\Msun\ (vdB+12) but a range $[5 - 25]\,10^{9}$. 

We now briefly comment on the mass ratio between the black hole and spheroidal 
component of the galaxy. \cite{vdB+12} find that a 17\,$10^9$~\Msun\ black hole would 
represent 59\% of the central bulge-like component. However, this directly relies on the definition
of what should be the "bulge" fraction of the galaxy (central or global component). By 
focusing on a specific scale, and thus a presumed central component, 
there is a risk that such a fraction is artificially set on the high side.
This again depends on a given scale and the (difficult) identification of what we 
should call a bulge. Since there is a direct connection between a given $M/L$ and
the corresponding best fit \MBH\ (see supplemental Figures 1 and 3 of vdB+12), it
obviously also depends on the $M/L$ itself (and the fraction of dark matter in the outer region), 
which directly influences the bulge mass. To further quantify this using our MGE model 
(with $M/L = 10$), and assuming we now identify the spheroidal component to
be the sum of all Gaussians with axis ratios larger than 0.3, we find a mass of $\sim1.8\, 10^{11}$~\Msun.
If we focus only on the more central spheroidal components (within e.g., 2~kpc) this would amount
to about $\sim 10^{11}$~\Msun. The large value advocated by \cite{vdB+12} for the 
black hole would thus represent between 10\% and 20\% of these masses, 
already significantly lower than the mentioned value of 59\%. Such an estimate would not be
significantly impacted by the presence of about 15\% of dark matter within one effective radius.
With a lower \MBH\ of 5\,$10^9$~\Msun\ as for Model~B of the present paper, 
this would correspond to either 3\% or 5\%, putting it in a range similar to what 
was derived for e.g., NGC\,4342 \cite{CrettonvdB99}. 
With a measured $\sigma$ of 333~\kms\ \cite{vdB+12},  the latter value for \MBH\ would make the 
black hole in NGC\,1277 only twice bigger than what a recent version of the 
\MBH-$\sigma$ relation predicts \cite{McConnell+11}, well within the measured scatter. 
NGC\,1277 would then again resemble a scaled-up version of e.g., NGC\,4342, another close to edge-on
fast rotator with a relatively large black hole: the fact that these are two, close to edge-on, 
fast rotators is an interesting fact to emphasise. In both cases, the spheroidal component contributes
significantly to the overal rotational support of the galaxy, with a relatively small tangential
anisotropy (at least near the equatorial plane). 

We have also presented an illustration of the impact of an inner bar in such models (Model~C and D) and
showed that the predicted central dispersion value could get relatively close (within 8\%) to the observed value
when such a mini-bar is viewed close to end-on, even without a central dark mass (the model fitting the dispersion
peak when \MBH$=2.5\,10^9$~\Msun). 
Such models naturally reproduce the relatively high $h_3$ amplitude, but fail to 
reproduce the observed positive central $h_4$ value.
We cannot comment further at this stage about the
possibility of fitting all observed kinematics simultaneously with a bar, as this would require more finely tuned
models to understand whether or not the central $h_4$ value is obtainable for
such systems. It is, however, probable that a central mass (of a few $10^9$~\Msun) would anyway be required if
the high positive $h_4$ value is confirmed, as it corresponds to broad wings 
in the central LOSVDs, thus stars moving at speeds significantly above
the predicted circular velocity (see Fig.~\ref{fig:Vcirc}).

This work emphasises the fact that the high value claimed for the mass of the central dark mass,
as quoted in \cite{vdB+12}, therefore mostly relies on the
robustness of the central (significantly positive) $h_4$ value. 
This seems quite natural considering that a high $h_4$ can be the direct consequence of
e.g., a high central mass concentration. The differences (in the high velocity wings of the LOSVDs) 
between our model with a 5~$10^9$~\Msun\ black hole and the measured central $h_4$ values cumulate
to about 5\,$10^8$ of stars rotating at about 1000~\kms. According to the predicted circular velocity profiles
shown in Fig.~\ref{fig:Vcirc}, this could correspond to a concentration of high angular momentum stars 
within the central 70 and 25~pc for \MBH\ of 17 and 5\,$10^9$~\Msun,
respectively: in the latter case, such a structure would not be resolved but should be detectable with
integral-field spectroscopy and adaptive optics or obtained in excellent seeing conditions.
Predictions from Model~B and~D for integral-field kinematics of the central few arcseconds at high resolution
(Full Width at Half Maximum, FWHM, of $0\farcs3$) are provided in Figs.~\ref{fig:ifu} and \ref{fig:barifu}:
the shape of the central stellar velocity dispersion peak (at that resolution), as well as the higher order moments 
should be strong discriminants between these various models as well as to further constrain the black hole mass.
Black holes of 2.5\,$10^9$ and 5\,$10^9$~\Msun would imply a very peaked central dispersion 
reaching about 445 and 500\,\kms\, respectively, at that resolution. 
We should expect a significantly more massive black hole as advocated by vdB+12 to produce an even more
extreme dispersion peak, reaching well beyond 500\,\kms. A central bar may show up a distinct kink in the 
$h_3$ map within 2\arcsec\ with local minima along the major-axis (see Fig.~\ref{fig:barifu}).

It may be difficult to rule out a lower mass black hole
using potential ionised or molecular gas associated with the dust ring, as this ring has
a diameter close to 1\arcsec\ where the circular velocity profiles of the various models (see Fig.~\ref{fig:Vcirc}) 
overlap, so that any interpretation would depend on the detailed characteristics of the gaseous orbits
(circular or not).
A proper test would again require two-dimensional spectroscopy  \cite[see e.g.][]{Krajnovic+05} at high spatial resolution
to probe the central few tenth of arcseconds\footnote{Gemini/NIFS data has now been obtained (PI: Richstone), 
and should help tremendously in this context.}. Another way
to start discriminating between these models would be if ionised gas is detected at {\em larger} radii (beyond a
few arcseconds) where the circular velocity profile is assumed to be dominated by the stellar component and where the
actual mass-to-light-ratio would have strong predictive power. 

Finally, we would like to emphasise the resemblance between a galaxy like NGC\,1277 and NGC\,4342, which
had been studied by \cite{CrettonvdB99} who claimed for the presence of supermassive black hole which represents
about 3\% of the total spheroid mass, one of the largest black hole mass to bulge mass known at
the time. NGC\,4342 is a close to edge-on fast rotator with a prominent 
inner disk and stellar kinematics which show similar radial major-axis profiles (though with different amplitudes),
specifically with an extended dispersion profile \citep{vdB+98}. The derived best fit model has anisotropy profiles 
\citep[see Figure~13 of][]{CrettonvdB99} which are 
very similar to the ones derived for NGC\,1277 here (Fig.~\ref{fig:Sigma}) with
$\sigma_{\theta}$ being lower in the outer part than the two other components of the velocity ellipsoid,
while going up in the central region above $\sigma_R$, and $\sigma_R / \sigma_z$ being rather constant
around 0.9. It is not surprising to note that, according to the model of NGC\,4342 
presented in \cite{CrettonvdB99}, the rise of $\sigma_{\theta}$ in the inner part 
corresponds to the observed rise in the stellar velocity dispersion. \cite{CrettonvdB99} also noted that the
mass-to-light ratio derived from their best fit model is large relatively to other early-type galaxies.
NGC\,1277 can thus roughly appear as a scaled-up version of NGC\,4342 (factor of about 8 in mass).
Interestingly, the potential presence of an end-on bar was discussed for NGC\,4342 by \cite{CrettonvdB99},
to acknowledge the similarities with another fast rotator, NGC\,4570, where evidence for bar-driven 
secular evolution \citep{vdBE98} was found. \cite{CrettonvdB99} considered the hypothesis of an end-on
bar in NGC\,4342 unlikely, reflecting on the fact that the end-on view would be rather constrained, 
and because the cuspy luminosity profile of NGC\,4342 would probably prevent
the formation of a bar in the first place (see the brief discussion in their Sect.~7.2.2). 
These two arguments could also apply to NGC\,1277 and NGC\,4570, while it is worth 
emphasising that the latter is also very cuspy and there are evidences
for NGC\,4570 to have been shaped by the presence of a bar.
One common denominator of both NGC\,4342 and NGC\,4570 is the
presence of a thin disk at a scale of about 100~pc, and NGC\,4570 harbours a relatively young ($\sim 2$~Gyr
old) and very thin stellar ring at a radius of 140~pc. In NGC\,1277, there is a clear ring-like dust structure
at a radius of about $0\farcs9$ ($\sim 320$~pc, at a distance of 73~Mpc), blocking part of the view
of the very central structure. We cannot comment further to whether or not 
these structures are various stages of an evolution process \cite[see also the 
beautiful and puzzling case of NGC\,1332, presented in][]{Rusli+11}, but
while searching for massive black holes in fast rotators with similar morphologies (and inclination),
it would be worth keeping these facts in mind.

\section*{Acknowledgements}
I would like to warmly thank Davor Krajnovi\'c for his helpful remarks on an early version of
the manuscript, as well as Michele Cappellari, and Glenn van de Ven for their input. 
I would like to specifically acknowledge very valuable feedback from Remco van den Bosch, including on the details of the 
Schwarzschild models presented in vdM+12, and thank him for the open and interesting discussions we had.
I wish to thank the anonymous referee for very constructive comments.
\bibliographystyle{mn2e}
\bibliography{NGC1277_MGE_astroph.bbl}

\end{document}